\documentclass[twocolumn,prl,aps,superscriptaddress,amsfonts,showpacs]{revtex4-1}

\usepackage{amsmath}
\usepackage{amssymb}
\usepackage{bm}
\usepackage{graphicx,psfrag}
\usepackage{epsfig}
\usepackage{afterpage}
\usepackage{mathrsfs}
\newcommand{\bwt}{\begin{widetext}}
\newcommand{\ewt}{\end{widetext}}
\newcommand{\bea}{\begin{eqnarray}}
\newcommand{\eea}{\end{eqnarray}}

\newcommand{\bef}{\begin{figure}[hbt]\centering}
\newcommand{\eef}{\end{figure}}

\begin{document}
\title{Testing the process dependence of the Sivers function \\
via hadron distributions inside a jet}

\date{\today}

\author{Umberto D'Alesio}
\affiliation{Dipartimento di Fisica,
             Universit\`a di Cagliari, Cittadella Universitaria,
             I-09042 Monserrato (CA), Italy}
\affiliation{Istituto Nazionale di Fisica Nucleare,
             Sezione di Cagliari, C.P. 170,
             I-09042 Monserrato (CA), Italy}

\author{Leonard Gamberg}
\affiliation{Division of Science,
             Penn State Berks,
             Reading, Pennsylvania 19610, USA}

\author{Zhong-Bo Kang}
\affiliation{RIKEN BNL Research Center,
             Brookhaven National Laboratory,
             Upton, New York 11973, USA}

\author{Francesco Murgia}
\affiliation{Istituto Nazionale di Fisica Nucleare,
              Sezione di Cagliari, C.P. 170,
              I-09042 Monserrato (CA), Italy}

\author{Cristian Pisano}
\affiliation{Dipartimento di Fisica,
             Universit\`a di Cagliari, Cittadella Universitaria,
             I-09042 Monserrato (CA), Italy}
 \affiliation{Istituto Nazionale di Fisica Nucleare,
              Sezione di Cagliari, C.P. 170,
              I-09042 Monserrato (CA), Italy}

\begin{abstract}
We study the process dependence of the Sivers function by considering the impact of color-gauge invariant initial and final state interactions on transverse spin asymmetries in proton-proton scattering reactions within the framework of the transverse momentum dependent (TMD), generalized parton model. To this aim, we consider the azimuthal distribution of leading pions inside a  fragmenting jet as well as single inclusive jet asymmetry in polarized proton-proton collisions.
In contrast to single inclusive pion production, in both cases we can isolate the Sivers contribution and thereby study its process dependence. The predictions for the Sivers asymmetry obtained with and without inclusion of color gauge factors are comparable in size but with opposite signs. We conclude that both processes represent unique opportunities to discriminate among the two approaches and test the universality properties of the Sivers function in hadronic scattering reactions.
\end{abstract}

\maketitle

Single transverse-spin asymmetries (SSAs) in high energy lepton-hadron and hadronic scattering processes have garnered  considerable attention from  both experimental and theoretical communities~\cite{D'Alesio:2007jt}.
Generally, they are defined as the ratio of the difference and the sum of the cross sections when the hadron's spin vector $S_\perp$ is flipped, $A_N\equiv (\sigma(S_\perp)-\sigma(-S_\perp))/(\sigma(S_\perp)+\sigma(-S_\perp))\equiv \Delta\sigma/(2\sigma^{\rm unp})$.
The SSAs for single inclusive particle production in proton-proton scattering are among the earliest processes studied~\cite{Bunce:1976yb} and remain extremely challenging to explain in the context of perturbative quantum chromodynamics (QCD)~\cite{Kang:2011hk}.
The trend of large SSAs in the pioneering fixed target experiments has been observed over a wide range of energies and more recently at significantly larger center-of-mass energies in the proton-proton collision experiments at Relativistic Heavy Ion Collider (RHIC)~\cite{Adams:2003fx, Adler:2005in}.
Also, azimuthal and transverse-spin asymmetries have been observed in Drell-Yan (DY) processes~\cite{Zhu:2006gx}, in semi-inclusive deep inelastic scattering (SIDIS)~\cite{:2009ti, :2008dn} and in hadron pair production in $e^+\, e^-$ scattering~\cite{Abe:2005zx}.

{}From a theoretical perspective SSAs are characterized by the interference between helicity flip and non-flip scattering amplitudes with a relative color phase. Two approaches have been proposed in the framework of perturbative QCD to account for these effects.
On the one hand is the collinear factorization formalism at next-to-leading-power (twist-3) in the hard scale where SSAs are given by a convolution of universal non-perturbative quark-gluon-quark correlation functions and hard scattering amplitudes~\cite{Efremov:1984ip, Qiu:1991pp, Kouvaris:2006zy}.

The other framework relies on factorization in terms of a hard scattering cross section and transverse momentum dependent (TMD) parton distribution and fragmentation functions (PDFs and FFs).
Prominent examples are the quark Sivers function~\cite{Sivers:1989cc}, which represents the azimuthal distribution of unpolarized quarks in a transversely polarized nucleon and the  Collins fragmentation function~\cite{Collins:1992kk}, which describes the production of pseudo-scalar mesons (or unpolarized hadrons) from transversely polarized fragmenting quarks.
In this approach color phases are given by initial and/or final state interactions (ISIs/FSIs) between the active quark and spectator remnants in the full scattering amplitude.
The details of the ISIs and FSIs depend on the scattering process and for PDFs such as the Sivers function, these color phases are incorporated into the Wilson lines of the gauge invariant definition of TMD PDFs.
It is a fundamental prediction of QCD factorization that the form of the gauge link depends on the hard sub-process~\cite{Bacchetta:2005rm} indicating that  the Sivers function is {\it non-universal}~\cite{Collins:2002kn}.
The oft-discussed case is the difference between the FSIs in SIDIS and the ISIs in DY scattering which leads to the prediction of an opposite relative color factor~\cite{Collins:2002kn}.
Further,  applying similar reasoning to hadron production in proton-proton collisions, typically the Sivers function has a more complicated color factor structure since both ISIs and FSIs contribute~\cite{Bacchetta:2005rm, Bacchetta:2007sz,Collins:2002kn,Gamberg:2010tj}.

While TMD factorization has not been established for hadron production in hadronic reactions~\cite{Collins:2007nk}, an extensive program of phenomenology has been carried out by including the correlations of intrinsic parton motion and transverse spin in the context of the so-called generalized parton model (GPM).
Introduced~\cite{Field:1976ve} as a generalization of the collinear perturbative QCD approach, it has been used to describe the SSAs for inclusive particle production~\cite{Anselmino:1994tv}.
Here factorization has been assumed as a reasonable starting point for analyses. At the same time, the leading-twist naive time-reversal odd (T-odd) TMD PDFs have  conditionally been  assumed to be {\it universal}.

In this Letter we present an analysis of SSAs in proton-proton scattering while taking into account the effects of ISIs and FSIs and allowing for process dependence within the framework of GPM. This will be referred to as the color gauge invariant (CGI) GPM. Previous studies along these lines have been carried out in~\cite{Bacchetta:2005rm, Bacchetta:2007sz,Collins:2002kn,Gamberg:2010tj, Gamberg:2010tj}.

We concentrate on reactions where one can explore the crucial issue of process dependence and universality of the Sivers function. Since several competing mechanisms can play a role in hadron collisions, following~\cite{D'Alesio:2010am} we consider the process $p^{\uparrow} p\rightarrow \, {\rm jet}\, \pi + X$, where one observes a large $p_T$ jet and looks for the azimuthal distribution of leading pions within the jet: a process under active investigation by the STAR Collaboration at RHIC~\cite{Fersch:2011zz}.
By contrast, an analysis where transverse partonic motion was considered only in the fragmentation process, aimed at a study of the universality of the Collins effect, was presented in~\cite{Yuan:2007nd}.  It is also important to note that the process we are studying is different from the case where {\it two} almost back-to-back hadrons or jets are observed as in~\cite{Bacchetta:2005rm,Bacchetta:2007sz}.  A {\it single} jet is measured in our study. Thus, the analysis of process dependence follows that carried out in~\cite{Gamberg:2010tj}.

Compared with inclusive pion production, we emphasize that the Sivers and Collins contributions can be disentangled in $p^{\uparrow} p\rightarrow \, {\rm jet}\, \pi + X$. Accordingly, in the GPM (keeping only the leading contributions after integration over the initial intrinsic transverse momenta~\cite{D'Alesio:2010am}) the numerator of the SSA for inclusive production of leading pions inside a large $p_{T}$ jet can be written as
\bwt
\bea
\frac{E_{\rm j}\, d\Delta\sigma}{d^3 {\bm p}_{\rm j} dz d^2 {\bm k}_{\perp \pi}}
& = & \frac{2\alpha_s^2}{s} \sum_{a,b,c,d} \int \frac{dx_a}{x_a}d^2\bm{k}_{\perp a}
\int \frac{dx_b}{x_b}d^2{\bm k}_{\perp b} \, \delta(\hat s+\hat t+\hat u)\,
H^U_{ab\to cd}(\hat s,\hat t,\hat u) \nonumber \\
&\times&\Big [ -\frac{k_{\perp a}}{M} f_{1T}^{\perp a}(x_a, {\bm k}_{\perp a}^2) \cos\phi_a
\, f_{b/B}(x_b, {\bm k}_{\perp b}^2)\,
D_{1}^c(z, {\bm k}_{\perp \pi}^2)
\sin\phi_{S_A} \nonumber\\
& & \mbox{}  + h_{1}^a(x_a, {\bm k}_{\perp a}^2) \cos(\phi_a-\psi)\, f_{b/B}(x_b, {\bm k}_{\perp b}^2)
\frac{k_{\perp \pi}}{z M_\pi}H_{1}^{\perp c}(z, {\bm k}_{\perp \pi}^2)
\,d_{NN}(\hat s,\hat t,\hat u) \,
\sin(\phi_{S_A}-\phi_\pi^H)\Big ]\,,
\label{main}
\eea
\ewt
where $E_{\rm j}$ and $\bm{p}_{\rm j}$ are the energy and momentum of the observed jet, $x_{a,b}$
and $\bm{k}_{\perp a,b}= k_{\perp a,b}(\cos\phi_{a,b}, \sin\phi_{a,b},0)$ are the initial parton light-cone momentum fractions and intrinsic transverse momenta respectively, and $z$ and $\bm{k}_{\perp\pi}$
($k_{\perp \pi}= |\bm{k}_{\perp\pi}|$) are the light-cone momentum fraction and transverse momentum of
the pion inside the jet with respect to the jet (parton $c$) direction of motion.

The second line of Eq.~(\ref{main}) corresponds to the Sivers effect, with the azimuthal modulation $\sin\phi_{S_A}$, where $\phi_{S_{A}}$ is the angle of the transverse spin vector, $S_A$, of hadron $A$ (with mass $M$), relative to the jet production plane and $f_{1T}^{\perp a}(x_a, \bm{k}_{\perp a}^2)$ is the Sivers function.
$H^U_{ab\to cd}(\hat s,\hat t, \hat u)$ is the unpolarized squared hard scattering amplitude for the process  $a\, b\to c\, d$, with $\hat s, \hat t, \hat u$ the usual partonic Mandelstam variables and $D_{1}^c(z,\bm{k}_{\perp\pi}^2)$ is the unintegrated fragmentation function for parton $c$ to fragment into a pion (with mass $M_\pi$). The third line of Eq.~(\ref{main}) corresponds to the Collins effect, with the azimuthal modulation $\sin(\phi_{S_A}-\phi_\pi^H)$, where $\phi_\pi^H$ is the azimuthal angle of the pion three-momentum around the jet thrust axis, as measured in the fragmenting parton helicity frame, and $H_{1}^{\perp c}(z,\bm{k}_{\perp\pi}^2)$ is the Collins function. It is convoluted with the unintegrated transversity distribution, $h_1^a(x_a, \bm{k}_{\perp a}^2)$, i.e.~the distribution of transversely polarized quarks in a transversely polarized hadron.
$d_{NN}$ is the partonic spin transfer asymmetry for the process $a^\uparrow b\to c^\uparrow d$ (defined as $(\sigma^{a^\uparrow b\to c^\uparrow d} - \sigma^{a^\uparrow b\to c^\downarrow d})/ (\sigma^{a^\uparrow b\to c^\uparrow d} + \sigma^{a^\uparrow b\to c^\downarrow d})$) and $\psi$ its azimuthal phase (for details see~\cite{D'Alesio:2010am}).

In close analogy with the case of SIDIS, one can define azimuthal moments and project out the various angular modulations in terms of $\phi_{S_A}$ and $\phi_\pi^H$ from  Eq.~(\ref{main}):
\bea
A_N^{W(\phi_{S_A},\phi_\pi^H)}(\bm{p}_{\rm j},z,k_{\perp\pi})=
\frac{\int d\phi_{S_A} d\phi_\pi^H\,
W(\phi_{S_A},\phi_\pi^H)\,d\Delta\sigma}
{\int d\phi_{S_A} d\phi_\pi^H\,
d\sigma^{\rm unp}}\,,
\nonumber
\eea
where $d\Delta\sigma=[d\sigma(\phi_{S_A},\phi_\pi^H) - d\sigma(\phi_{S_A}+\pi,\phi_\pi^H)]$, as given in Eq.~(\ref{main}) and $d\sigma^{\rm unp}=[d\sigma(\phi_{S_A},\phi_\pi^H) + d\sigma(\phi_{S_A}+\pi,\phi_\pi^H)]/2$ is the unpolarized cross section. 
By choosing $W(\phi_{S_A},\phi_\pi^H)=\sin\phi_{S_{A}}$ one then singles out the Sivers contribution to $A_N$, that is $A_N^{\sin\phi_{S_A}}$, which we focus on in the following.  Moreover, since our aim is to study the process dependence of the {\em quark} Sivers function, we will consider pion-jet production at large rapidities. Here, any potential sea-quark and gluon Sivers effects are expected to be negligible, as follows from the analyses of SSAs in SIDIS (see \cite{Anselmino:2008sga}) and in $pp\to \pi + X$ at midrapidity~\cite{Anselmino:2006yq, Adler:2005in, Wei:2011nt} and from 
the study carried out in~\cite{Brodsky:2006ha}.

\begin{figure*}[hbt]
\includegraphics[width=2.2in]{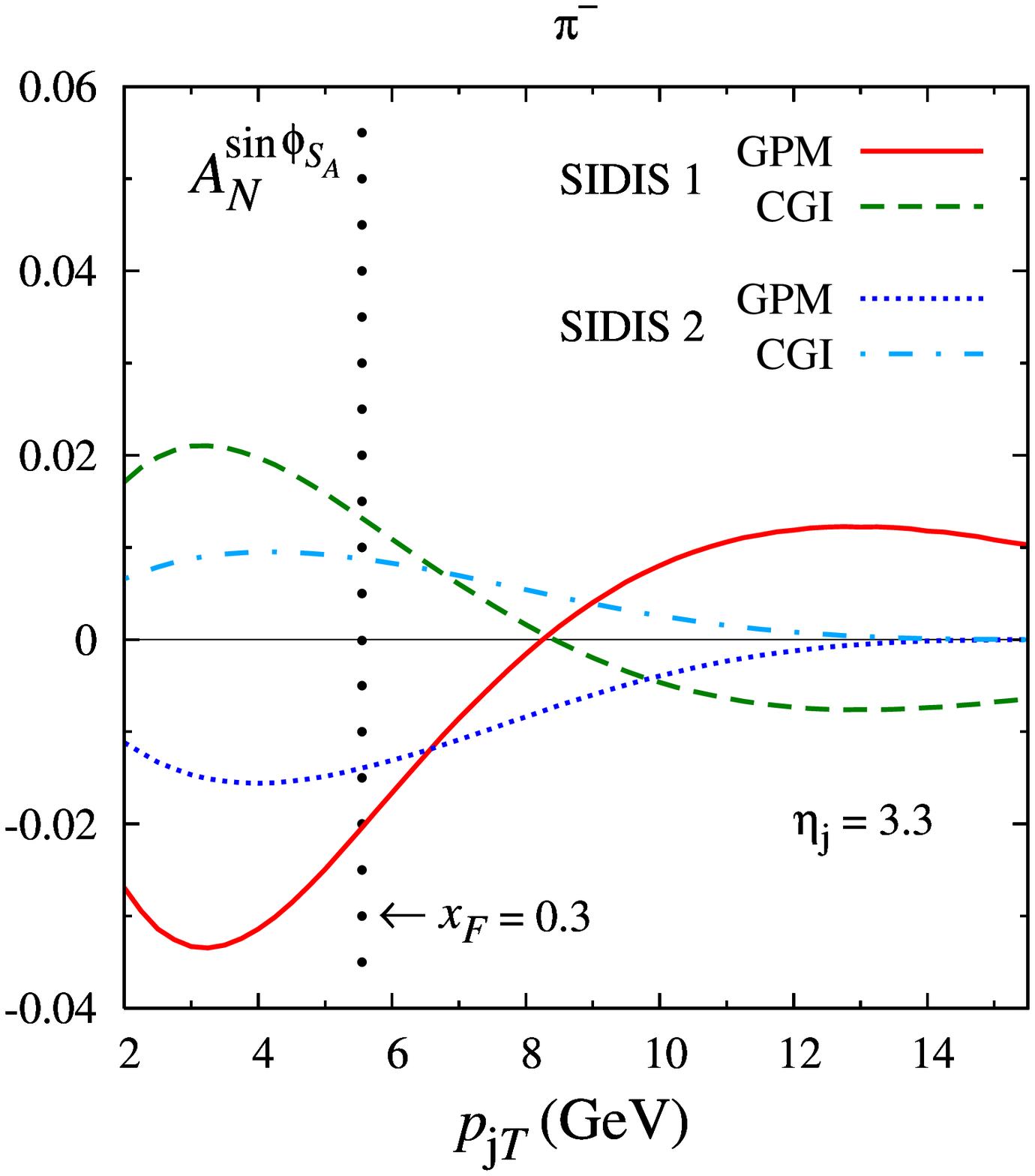}
\includegraphics[width=2.2in]{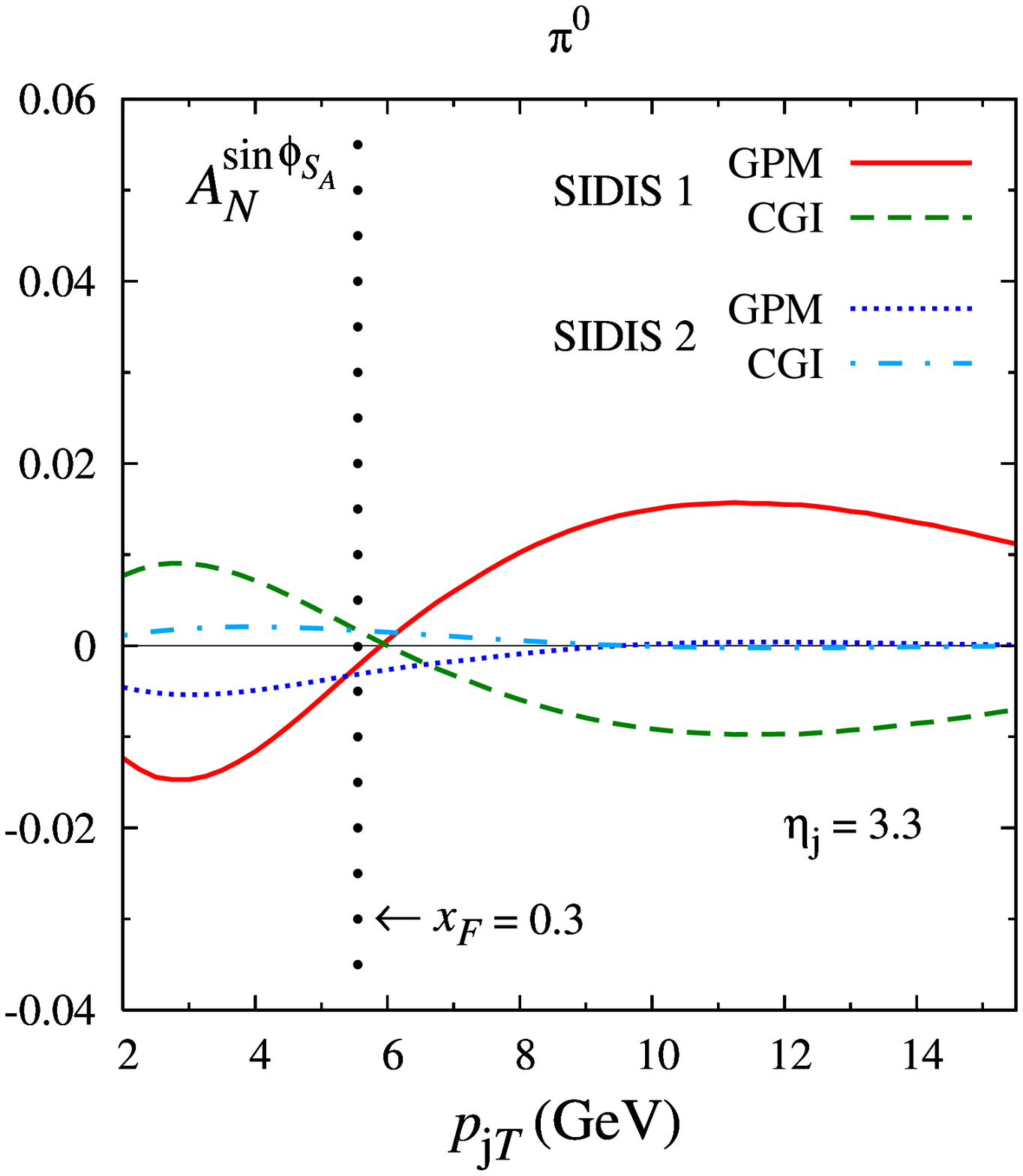}
\includegraphics[width=2.2in]{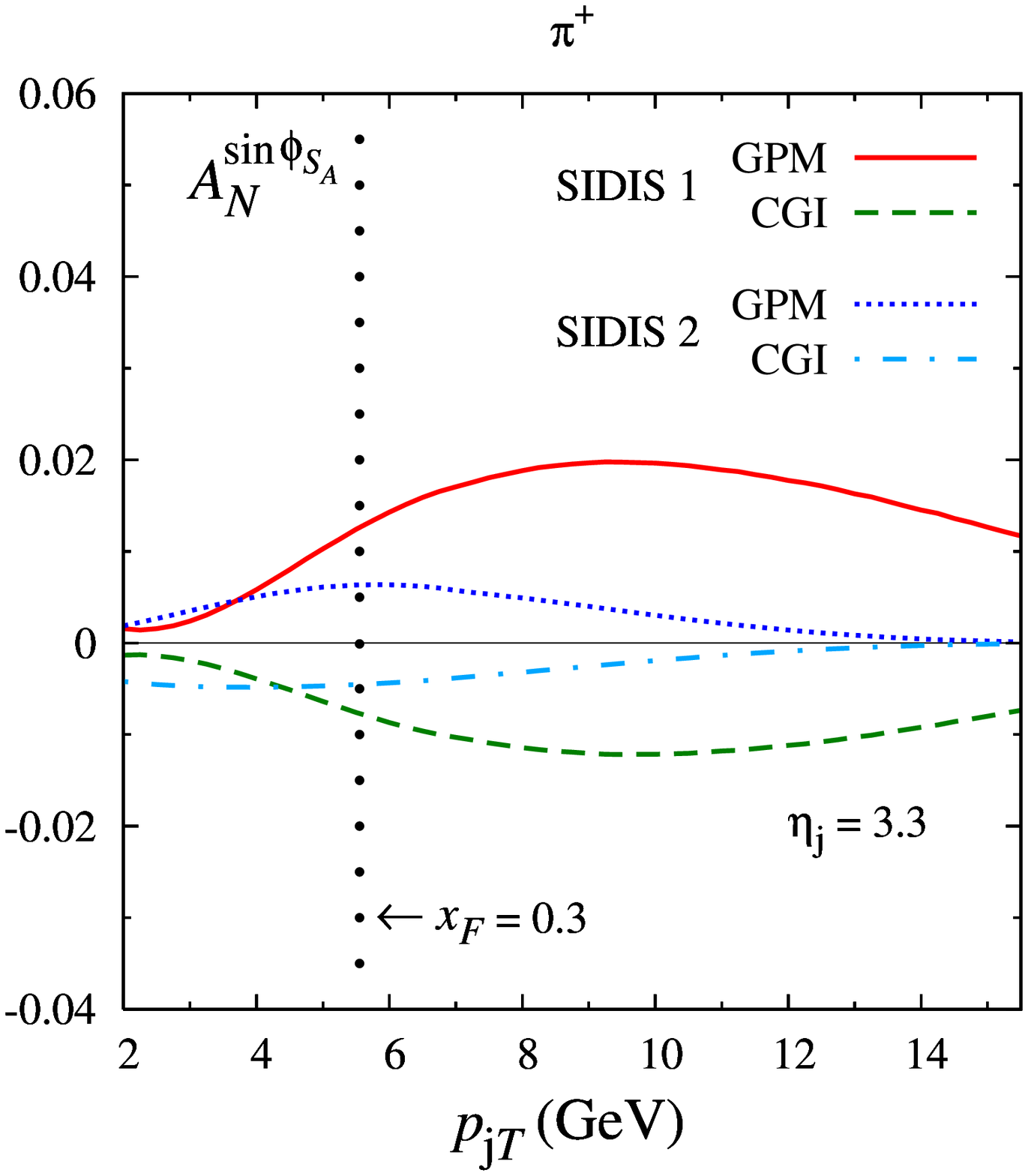}
\caption{The Sivers asymmetry $A_N^{\sin\phi_{S_A}}$ for the $p^\uparrow p\to {\rm jet}\, \pi +X$
as a function of $p_{{\rm j}T}$, at fixed jet rapidity $\eta_{\rm j}=3.3$, for RHIC energy, $\sqrt{s}=500$ GeV. The solid (SIDIS~1~\cite{Anselmino:2005ea}) and dotted (SIDIS~2~\cite{Anselmino:2008sga}) curves
are for the GPM calculation, and the dashed and dot-dashed ones for the CGI GPM calculation. The vertical dotted line corresponds to $x_F=0.3$.}
\label{fig1}
\end{figure*}

In the GPM, the Sivers function is assumed to be universal and taken to be the same as that probed in SIDIS; that is in Eq.~(\ref{main}),
\bea
f_{1T}^{\perp a}(x_a, \bm{k}_{\perp a}^2) \equiv f_{1T}^{\perp a, \rm SIDIS}(x_a, \bm{k}_{\perp a}^2).
\eea
On the other hand, for the process $p^\uparrow p\to {\rm jet}\, \pi +X$, both ISIs and FSIs contribute and thus in principle the Sivers function for the pion-jet production should be different from that probed in SIDIS.
Following~\cite{Gamberg:2010tj}, we carefully analyze these ISIs and FSIs for all the partonic scattering processes relevant to the azimuthal distribution of leading pions inside a fragmenting jet in proton-proton scattering. In this way the (quark) Sivers contribution from the CGI GPM will be (here $a\equiv q$)
\bea
\frac{E_{\rm j}\, d\Delta\sigma^{(\rm Siv)}}{d^3 {\bm p}_{\rm j} dz d^2 {\bm k}_{\perp \pi}}
& = & \frac{2\, \alpha_s^2}{s} \sum_{a,b,c,d} \int \frac{dx_a}{x_a}d^2\bm{k}_{\perp a}
\int \frac{dx_b}{x_b}d^2{\bm k}_{\perp b}
\nonumber\\
&&\times
\,\delta(\hat s+\hat t+\hat u)\,H^{U}_{ab\to cd}(\hat s,\hat t,\hat u)
\nonumber \\
&&\times
\Big ( -\frac{k_{\perp a}}{M}\Big)f_{1T}^{\perp a, ab\to cd}(x_a, {\bm k}_{\perp a}^2) \cos\phi_a
\nonumber \\
&&\times
\, f_{b/B}(x_b, {\bm k}_{\perp b}^2)\, D_{1}^c(z, {\bm k}_{\perp \pi}^2) \sin\phi_{S_A} \,,
\label{process}
\eea
in which a {\it process-dependent} Sivers function denoted as $f_{1T}^{\perp a, ab\to cd}$ is used rather than that from SIDIS as in the GPM approach~\cite{Anselmino:1994tv, D'Alesio:2010am}.
The crucial point is that the existence of the Sivers function in the polarized nucleon relies on the ISIs and FSIs between the struck parton and the spectators from the polarized nucleon through the gluon exchange. Thus by analyzing these interactions, one can compute the color factors $C_I$ ($C_{F_c}$) for initial (final) state interactions that determine the process dependent Sivers function to be used for the corresponding partonic scattering $a\, b\to c\, d$. In the CGI GPM, the process dependence of the Sivers function can be shifted to the squared hard partonic scattering amplitude, that is
\begin{equation}
f_{1T}^{\perp a, ab\to cd} \,H^{U}_{ab\to cd} \equiv f_{1T}^{\perp a, \rm SIDIS} \,H^{\rm Inc}_{ab\to cd}\,,
\end{equation}
where all the process dependence is absorbed into the new hard function $H^{\rm Inc}_{ab\to cd}$, which is the same as in the single inclusive particle production~\cite{Gamberg:2010tj}. This approach suggests a close connection with the twist-3 collinear  formalism~\cite{Qiu:1991pp, Kouvaris:2006zy} (see~\cite{Gamberg:2010tj} for details).

Now we study the consequence of these ISIs and FSIs by comparing the predictions of the Sivers
asymmetry for pion-jet production between GPM and CGI GPM. In Fig.~\ref{fig1} we plot $A_N^{\sin\phi_
{S_A}}(\bm{p}_{\rm j})$ for $\pi^{0,\pm}$-jet production, as a function of the jet transverse momentum
$p_{{\rm j}T}$ at forward rapidity, $\eta_{\rm j}=3.3$, for RHIC energy, $\sqrt{s}=500$ GeV, integrated over $\bm{k}_{\perp\pi}$ and $z$ ($z\ge 0.3$)~\cite{D'Alesio:2010am}.
The estimates using the two available parameterizations of the Sivers function in the GPM formalism are shown as the solid (SIDIS~1~\cite{Anselmino:2005ea}) and dotted (SIDIS~2~\cite{Anselmino:2008sga}) lines, while the corresponding ones using CGI GPM formalism in Eq.~(\ref{process}) are shown as dashed and
dot-dashed lines. One immediately sees that the results of the two approaches, while comparable in
size, exhibit {\it different} signs. The opposite sign is the manifestation of ISIs and FSIs, see Fig.~\ref{fig2} for illustration. Particularly for the dominant channel at forward rapidity, $qg\to qg$ with the final quark identified with the observed jet, these ISIs/FSIs lead to $H^{\rm Inc}_{qg\to qg}\sim -\frac {N_c^2+2}{N_c^2-1}\frac{\hat s^2}{\hat t^2}$ for CGI GPM, while $H^{U}_{qg\to qg}\sim \frac{2\hat s^2} {\hat t^2}$ for GPM~\cite{Gamberg:2010tj}.

\bef
\psfig{file=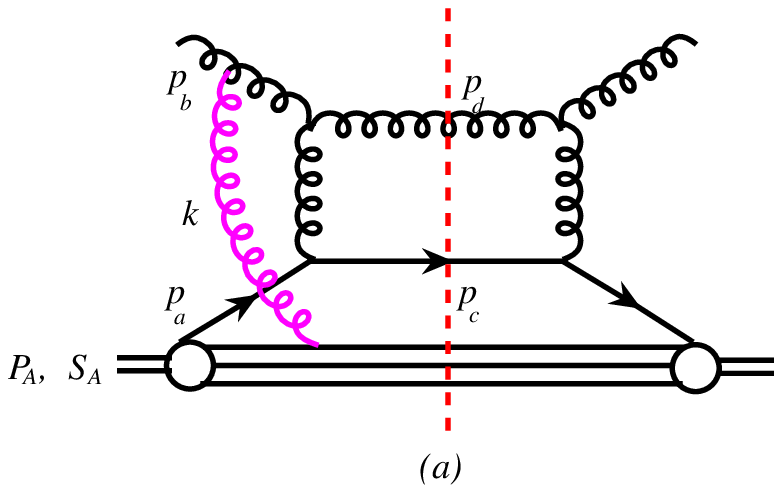, width=1.5in}
\psfig{file=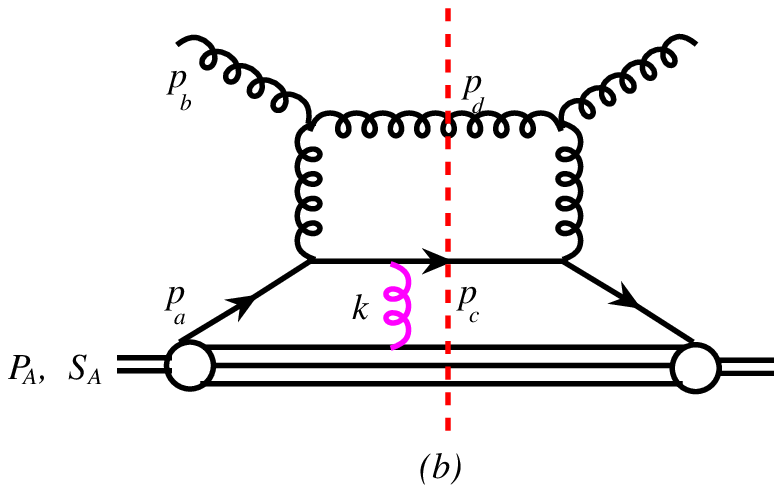, width=1.5in}
\caption{Sample Feynman diagrams for the initial (a) and final (b) state interactions for pion-jet production, illustrated here for the partonic channel $qg\to qg$ with the final $q$ identified with the observed jet.}
\label{fig2}
\eef

The predictions labeled SIDIS~1 and SIDIS~2 are similar in the intermediate $p_{{\rm j}T}\lesssim 5.5$ GeV region (corresponding to Feynman $x$, $x_F<0.3$), where the Sivers function parameterizations are constrained by present SIDIS data (that is at Bjorken $x$, $x_B<0.3$). This region is then optimal to test directly the process dependence of the Sivers function. This is the main goal of our analysis. Moreover, the observation of a sizable $A_N^{\sin\phi_{S_A}}$ at large $p_{{\rm j}T}$ (i.e.~large $x_F$) could be extremely useful to constrain the Sivers function in the large $x$ region~\cite{D'Alesio:2010am}, as well as to test its process dependence. In this respect large $x_B$-data from SIDIS~\cite{Qian:2011py}, e.g., at Jefferson Lab in the 12 GeV program, could be very important.

Note that for $\sqrt s = 200$ GeV the behavior of our estimates would be similar to that shown in Fig.~\ref{fig1}, gaining almost a factor of 2 in size. However the range of $p_{{\rm j}T}$ covered would be narrower ($p_{{\rm j}T}\le 6.5$~GeV) and with $x_F \le 0.3$ now corresponding to $p_{{\rm j}T}\le 2.2$~GeV.

As a natural extension of this work we can consider single inclusive jet asymmetry in proton-proton scattering by replacing the fragmentation function $D_{1}^c(z,\bm{k}^2_{\perp \pi})$ in
Eqs.~(\ref{main}) and (\ref{process}), by $\delta(z-1)\,\delta^2(\bm{k}_{\perp \pi})$. In this case the SSAs are described solely by the Sivers function. An analogous study of the Sivers contribution yields a similar process dependence and the results we obtain for $A_N^{\sin\phi_{S_A}}$ (not shown) look almost indistinguishable from the case of neutral pion-jet production (central panel of Fig.~\ref{fig1}).

In summary, we have studied the azimuthal distribution of leading pions inside a jet as well as single inclusive jet production in proton-proton scattering, under present active investigation at RHIC.
By adopting the TMD GPM, we have considered ISIs and FSIs leading to process dependence of the Sivers function.
We have presented estimates of the Sivers asymmetry $A_N^{\sin\phi_{S_A}}$ for RHIC kinematics within the GPM framework with and without inclusion of color gauge factors. We find that the resulting Sivers asymmetries are comparable in size but appear with opposite signs. We conclude that the experimental observation of a sizable $A_N^{\sin\phi_{S_A}}$ in $p^\uparrow p\to {\rm jet}\, \pi +X$ or $p^\uparrow p\to {\rm jet} + X$ can test the role color gauge invariance plays in the universality properties of the Sivers function. At the same time it could give a clean indication on the size of the Sivers function in the large $x$ region (not covered by present SIDIS data). This will definitely provide new insights into our understanding of single spin asymmetries in QCD.

\vskip 0.1in
We are grateful to M.~Anselmino for carefully reading the manuscript.
This work was supported in part by U.S. Department of Energy under Grant No. DE-FG02-07ER41460 (L.G.) and Contract No. DE-AC02-98CH10886 (Z.K.).
U.D.~and F.M.~acknowledge partial support by Italian MIUR under PRIN 2008,
and by the European Community under the FP7 grant agreement No.~227431.
C.P.~is supported by Regione Autonoma della Sardegna 
under grant PO Sardegna FSE 2007-2013, L.R. 7/2007.

\end{document}